\documentclass[preprintnumbers]{revtex4}
\pdfoutput=1

\usepackage{amsmath,bbm,latexsym,url}

\newcommand*{\fvec}[1]{\ensuremath{\boldsymbol{\mathrm{\tilde{#1}}}}  }    
\newcommand*{\fmat}[1]{\tilde{#1}}                   
\newcommand*{\tvec}[1]{\ensuremath{\boldsymbol{\mathrm{#1}}}}           
\newcommand*{\trans}{\mathrm{T}}                     
\DeclareMathOperator{\re}{Re}
\DeclareMathOperator{\im}{Im}
\DeclareMathOperator{\diag}{diag}

\numberwithin{equation}{section}

\begin{document}

\title{CP properties of symmetry-constrained two-Higgs-doublet models}

\author{P.M.~Ferreira,$^{(1,2)}$\thanks{E-mail: ferreira@cii.fc.ul.pt}
\\*[2mm]
M.~Maniatis,$^{(3)}$\thanks{E-mail: m.maniatis@thphys.uni-heidelberg.de}
\\*[2mm]
O.~Nachtmann,$^{(3)}$\thanks{E-mail: o.nachtmann@thphys.uni-heidelberg.de}
\\*[2mm]
and Jo\~{a}o P.~Silva$\, ^{(1,4)}$\thanks{E-mail: jpsilva@cftp.ist.utl.pt}
\\*[3mm]
\small $^{(1)}$ Instituto Superior de Engenharia de Lisboa \\
\small 1959-007 Lisboa, Portugal
\\*[2mm]
\small $^{(2)}$ Centro de F\'\i sica Te\'orica e Computacional,
Universidade de Lisboa \\
\small 1649-003 Lisboa, Portugal
\\*[2mm]
\small $^{(3)}$ Institut f\"ur Theoretische Physik \\
\small Philosophenweg 16, 69120 Heidelberg, Germany
\\*[2mm]
\small $^{(4)}$ Centro de F\'\i sica Te\'orica de Part\'\i culas,
Instituto Superior T\'ecnico \\
\small 1049-001 Lisboa, Portugal
}

\preprint{HD-THEP-10-8}

\begin{abstract}
The two-Higgs-doublet model can be constrained by imposing
Higgs-family symmetries and/or generalized CP symmetries.
It is known that there are only six independent classes
of such symmetry-constrained models.
We study the CP properties of all cases in
the bilinear formalism.
An exact symmetry implies CP conservation.
We show that soft breaking of the symmetry can lead to
spontaneous CP violation (CPV) in three of the classes.
\end{abstract}

\maketitle

\section{Introduction and Notation}

Experiments will soon probe the electroweak symmetry
breaking (EWSB) mechanism.
This is achieved in the Standard Model (SM) with
one Higgs doublet.
A very simple extension involves two Higgs-field doublets
$\varphi_1$ and $\varphi_2$,
required also by supersymmetry.
One particular feature of this model that makes it very popular is the
possibility that it has vacua which spontaneously break the CP
symmetry. Indeed this was the main reason why Lee introduced
the two-Higgs-doublet model (THDM) in 1973~\cite{Lee:1973iz}.
In fact, it is known that the amount of CP violation that the Standard Model
displays is insufficient to explain the observed matter-antimatter
asymmetry in the universe. As such, models wherein CP violation
arises from other mechanisms are in demand, and the THDM is the simplest
such model. 
We will investigate whether
CP is preserved in the potential, both before and after EWSB. 
We will consider all possible symmetry-constrained THDM
potentials, both with exact and softly broken symmetries. And we will
be able to conduct this study in an analytical manner, for most
cases avoiding direct calculation of the fields' vacuum expectation
values.

The most general potential of the THDM may be written in terms of fields as
\cite{Gunion:1989we,Gunion:1992hs}
\begin{eqnarray}
V &=&
m_{11}^2 (\varphi_1^\dagger \varphi_1) +
m_{22}^2 (\varphi_2^\dagger \varphi_2) -
m_{12}^2 (\varphi_1^\dagger \varphi_2) -
(m_{12}^2)^* (\varphi_2^\dagger \varphi_1)
\nonumber\\
&&
+\frac{1}{2} \lambda_1 (\varphi_1^\dagger \varphi_1)^2
+ \frac{1}{2} \lambda_2 (\varphi_2^\dagger \varphi_2)^2
+ \lambda_3 (\varphi_1^\dagger \varphi_1)(\varphi_2^\dagger \varphi_2)
\nonumber\\
&&
+ \lambda_4 (\varphi_1^\dagger \varphi_2)(\varphi_2^\dagger \varphi_1)
+ \frac{1}{2} [\lambda_5 (\varphi_1^\dagger \varphi_2)^2 + \lambda_5^*
(\varphi_2^\dagger \varphi_1)^2]
\nonumber\\
&&
+ [\lambda_6 (\varphi_1^\dagger \varphi_2) + \lambda_6^*
(\varphi_2^\dagger \varphi_1)] (\varphi_1^\dagger \varphi_1) + [\lambda_7 (\varphi_1^\dagger
\varphi_2) + \lambda_7^* (\varphi_2^\dagger \varphi_1)] (\varphi_2^\dagger \varphi_2)\;,
\label{V_fields}
\end{eqnarray}
with $m_{11}^2$, $m_{22}^2$, $\lambda_{1,2,3,4}$ real and
$m_{12}^2$, $\lambda_{5,6,7}$ complex.
To study the properties of the Higgs potential,
it is oftentimes easier to write it in terms of
field bilinears \cite{Velhinho:1994vh,Nagel:2004sw,Ferreira:2004yd, Ivanov:2005hg,Man1,Nishi:2006tg,Ivanov:2006yq,Man2,
Nishi:2007nh,Ivanov:2007de,Nishi:2007dv,Man3}.
In~\cite{Nagel:2004sw, Man1, Nishi:2006tg}
a one-to-one correspondence of
gauge invariant expressions with
bilinears with a simple
geometric interpretation was revealed. In this approach the bilinears
form a Minkowski type four vector. We follow the notation of the Heidelberg
group~\cite{Nagel:2004sw,Man1} and write~$V$ as
\begin{equation}
\label{VK}
V =
\fvec{K}^\trans \fvec{\xi} + \fvec{K}^\trans \fmat{E} \fvec{K}
\end{equation}
with
\begin{equation}
\fvec{K} =
\left(
\begin{array}{c}
K_0 \\
\tvec{K}
\end{array}
\right),
\ \ \
K_0 = \varphi_1^\dagger \varphi_1 + \varphi_2^\dagger \varphi_2,
\ \ \
\tvec{K} =
\left(
\begin{array}{c}
\varphi_1^\dagger \varphi_2 + \varphi_2^\dagger \varphi_1 \\
i \varphi_2^\dagger \varphi_1 -  i \varphi_1^\dagger \varphi_2 \\
\varphi_1^\dagger \varphi_1 - \varphi_2^\dagger \varphi_2
\end{array}
\right).
\label{connect_4}
\end{equation}
In~\eqref{VK}~$\fvec{\xi}$ and $\fmat{E}$ contain the parameters,
which are all real:
\begin{equation}
\fvec{\xi} =
\left(
\begin{array}{c}
\xi_0\\
\tvec{\xi}
\end{array}
\right),
\ \ \ \
\fmat{E} =
\left(
\begin{array}{cc}
\eta_{00} & \tvec{\eta}^\trans\\
\tvec{\eta} & E
\end{array}
\right),
\label{eq-fourpar}
\end{equation}
with $E=E^\trans$ a $3\times 3$~matrix.
Expressing $\tilde{\tvec{\xi}}$ and $\tilde{\tvec{E}}$ in terms
of the parameters of~\eqref{V_fields},
we find
\begin{eqnarray}
\xi_0=\frac{1}{2}
(m_{11}^2+m_{22}^2)\;,
&\hspace{5mm}&
\tvec{\xi}=\frac{1}{2}
\left(
\begin{array}{c}
- 2 \textrm{Re}(m_{12}^2)\\
2 \textrm{Im}(m_{12}^2)\\
 m_{11}^2-m_{22}^2
\end{array}
\right),
\label{connect_1}\\*[3mm]
\eta_{00} =
\frac{1}{8}(\lambda_1 + \lambda_2) + \frac{1}{4}\lambda_3\;,
&\hspace{5mm}&
\tvec{\eta}=\frac{1}{4}
\left(
\begin{array}{c}
\textrm{Re}(\lambda_6+\lambda_7)\\
-\textrm{Im}(\lambda_6+\lambda_7)\\
\frac{1}{2}(\lambda_1 - \lambda_2)
\end{array}
\right),
\label{connect_2}
\end{eqnarray}
\begin{equation}
E = \frac{1}{4}
\left(
\begin{array}{ccc}
\lambda_4 + \textrm{Re}(\lambda_5) &
-\textrm{Im}(\lambda_5) &
\textrm{Re}(\lambda_6-\lambda_7) \\
-\textrm{Im}(\lambda_5) &
\lambda_4 - \textrm{Re}(\lambda_5) &
-\textrm{Im}(\lambda_6-\lambda_7) \\
\textrm{Re}(\lambda_6-\lambda_7) &
-\textrm{Im}(\lambda_6 -\lambda_7) &
\frac{1}{2}(\lambda_1 + \lambda_2) - \lambda_3
\end{array}
\right),
\label{connect_3}
\end{equation}
When using bilinears one must only take care that,
by construction,
they are restricted by
\begin{equation}
K_0 \geq 0,\ \  \textrm{and}\ \  K_0^2 - \tvec{K}^2 \geq 0.
\label{restrictions}
\end{equation}\\

The fields $\varphi_1$ and $\varphi_2$ are not physical,
only their mass eigenstates will be physical.
Thus, we can choose for general discussions, for instance of symmetries and
their breaking,
any linear combination thereof that doesn't change the
kinetic terms. The physics will have to remain the same for these {\em basis changes},
i.e. unitary transformations of the form
\begin{equation}
\label{eq11}
\begin{pmatrix} \varphi_1 \\
                \varphi_2 \end{pmatrix}
\to
\begin{pmatrix} \varphi'_1 \\
                \varphi'_2 \end{pmatrix}
= U
  \begin{pmatrix} \varphi_1 \\
                  \varphi_2 \end{pmatrix}\;,
\end{equation}
with $U= (U_{ij}) \in U(2)$. In bilinear language, this corresponds
to a $SO(3)$ rotation in $K$ space, given by
\begin{equation}
\label{eq12}
\begin{split}
&K_0 \to K'_0 = K_0,\\
&\tvec{K} \to \tvec{K}' = R(U)  \tvec{K}\;.
\end{split}
\end{equation}
Here $R(U)$ is obtained from
\begin{equation}
\label{eq13}
U^\dagger \sigma^a U = R_{ab}(U)\,\sigma^b.
\end{equation}
This freedom of choosing a basis may help to simplify many calculations, but it also
may mask the symmetry of a potential. Thus, basis-invariant methods are
advantageous in many problems.\\

CP properties of the THDM Higgs potential were studied extensively
in the literature in a basis-independent
way~\cite{Gun, Lavoura:1994fv, Botella:1994cs,Ginzburg:2004vp, Branco:2005em, Davidson:2005cw, Gunion:2005ja, Nishi:2006tg,Man2,Nishi:2007nh,Nishi:2007dv}.
Here we focus on a study of CP properties in terms of the
bilinears. As shown in~\cite{Man2}, the standard CP transformation
of the Higgs doublets
\begin{equation}
\varphi_i(x) \rightarrow \varphi_i^*(x') , \qquad i=1,2
\end{equation}
corresponds in $K$ space to
\begin{equation}
\label{standCPK}
\begin{split}
K_0(x) &\rightarrow K_0(x')\\
\tvec{K}(x) &\rightarrow \diag(1,-1,1) \; \tvec{K}(x') .
\end{split}
\end{equation}
The parity transformation flips the sign of the spatial components
in $x \rightarrow x'$.
Geometrically, a standard CP transformation
is a reflection of the $\tvec{K}$ vector
on the 1--3 plane in a certain basis
in addition to the parity transformation
of the argument.

We shall also consider generalized CP transformations~(GCPs) as discussed already in~\cite{Lee:1966ik,Ecker:1981wv,Ecker:1983hz,Bernabeu:1986fc,Ecker:1987qp,Neufeld:1987wa,Lavoura:1994fv, Botella:1994cs}, where
the standard CP transformation
is accompanied by a mixing of the Higgs fields:
\begin{equation}
\label{eqX}
\varphi_i(x) \rightarrow X_{ij} \varphi_j^*(x') , \qquad i,j=1,2\; ,
\end{equation}
with $X=(X_{ij}) \in U(2)$.
In a suitable basis for the fields
the matrix $X$ in~\eqref{eqX} can
always be brought
to the form~\cite{Ecker:1987qp}
\begin{equation}
\label{eq32}
X=
\begin{pmatrix}
\cos (\theta) & \sin (\theta) \\
-\sin (\theta) & \cos (\theta)
\end{pmatrix},
\qquad \text{with } 0 \le \theta \le \pi/2 .
\end{equation}
These transformations were classified
in~\cite{Man2,Ferreira:2009wh}. Following here
the notation in~\cite{Ferreira:2009wh}
we have
\begin{equation}
\label{GCPclass}
\begin{split}
& \text{CP1} \quad \text{ if }\quad \theta =0 , \\
& \text{CP2} \quad \text{ if } \quad \theta = \pi/2 , \\
& \text{CP3} \quad \text{ if } \quad 0< \theta < \pi/2.
\end{split}
\end{equation}

In $K$ space the GCP transformations~\eqref{eqX} correspond 
to the improper rotations
\begin{equation}
\tvec{K}(x) \to \bar{R} \; \tvec{K}(x')
\end{equation}
with
\begin{equation}
\label{GCPK}
\bar{R}=
\begin{pmatrix}
\cos (2 \theta) & 0 & -\sin (2 \theta)\\
0 & -1 & 0\\
\sin (2 \theta) & 0 & \cos (2 \theta)
\end{pmatrix},
\qquad  0 \le \theta \le \pi/2
\end{equation}
in the specific basis where $X$ has the form~\eqref{eq32}.
From this representation we identify CP1 as reflections on planes and
CP2 as the point reflection at the origin in addition to the parity transformation
of the argument.\\

We emphasize that the symmetries CP1, CP2, and CP3 are independent symmetries, which
may be imposed on the Lagrangian. However, considering only the
THDM scalar sector, invariance under CP2 or CP3 implies invariance under
CP1, as shown in~\cite{Ferreira:2009wh}. This will also be seen
directly from the discussion in section~\ref{secmodels} below.
Furthermore, considering only the scalar sector, no
Higgs basis is preferred and, a priori,
any GCP transformation of type CP1 is as good
as any other to be called {\em the} CP transformation.
In the sequel, we will, therefore, simply
talk about {\em CP symmetry}, {\em CP conservation},
and {\em CP violation} and drop the notation generalized.\\

In the following, we will use basis-invariant quantities, that is,
quantities invariant under the transformations~\eqref{eq11},
respectively in $K$~space, the rotations~\eqref{eq12}. We shall
also work in bases which are clearly singled out by the problem at hand.
This can be the basis implied by~\eqref{eq32}, respectively \eqref{GCPK},
or the basis where $E$ is diagonal. Note that the
eigenvalues of $E$ and the angle $\theta$ in 
\eqref{eq32}, \eqref{GCPK} are basis independent quantities.
For the eigenvalues of~$E$ this is so by construction.
To see this for the angle~$\theta$ we note that by a basis
transformation~\eqref{eq11}~$X$ in \eqref{eqX} is replaced by
\begin{equation}
X'= U X U^\trans \;.
\end{equation}
This implies
\begin{equation}
\label{1.18b}
X' X'^* = U X X^* U^\dagger\;.
\end{equation}
It is easy to see from~\eqref{eq32} that $\exp (\pm 2 i \theta )$
are the eigenvalues of $X X^*$ in the standard form.
But using~\eqref{1.18b} we see that these are then also the
eigenvalues of $X X^*$ in any basis, that is, $\theta$ is basis
independent.\\

One can construct 
basis invariant quantities which measure explicit CP violation at the
level of the THDM Lagrangian, that is, before
EWSB~\cite{Gun, Ginzburg:2004vp, Branco:2005em, Davidson:2005cw, Gunion:2005ja, Nishi:2006tg,Man2}.
In terms of the potential parameters~\eqref{connect_1}--\eqref{connect_3}
a convenient set of such quantities (see theorem~3 of~\cite{Man2})
is given by~\footnote{Our convention for the $I's$ is equivalent to but
differs from those adopted in \cite{Gun, Ginzburg:2004vp, Branco:2005em, Davidson:2005cw, Gunion:2005ja, Nishi:2006tg}.}
\begin{eqnarray}
I_1 &=& \left(\tvec{\xi} \times \tvec{\eta} \right)^\trans\ E \tvec{\xi},
\label{I1}
\\
I_2 &=& \left(\tvec{\xi} \times \tvec{\eta} \right)^\trans\ E \tvec{\eta},
\label{I2}
\\
I_3 &=& \left(\tvec{\xi} \times (E \tvec{\xi}) \right)^\trans\ E^2 \tvec{\xi},
\label{I3}
\\
I_4 &=& \left(\tvec{\eta} \times (E \tvec{\eta}) \right)^\trans\ E^2 \tvec{\eta}.
\label{I4}
\end{eqnarray}
The theory is CP conserving at
the Lagrangian level, meaning that it
allows for {\em at least one} CP transformation as a symmetry,
if and only if all $I's$ vanish.

Suppose now that the Higgs potential is indeed invariant under
at least one CP transformation, that is, all $I$'s vanish.
Then, there is {\em at least one} CP invariance respected simultaneously at the
Lagrangian level {\em and} by the
vacuum if also the three following basis-invariant quantities
vanish~\footnote{Our convention for the $J's$ follows~\cite{Man2} and
is equivalent to but
differs from that adopted in \cite{Lavoura:1994fv,Botella:1994cs}.},
\begin{eqnarray}
J_1 &=& \left(\tvec{\xi} \times \tvec{\eta} \right)^\trans\ \langle \tvec{K} \rangle,
\label{J1}
\\
J_2 &=& \left(\tvec{\xi} \times (E \tvec{\xi}) \right)^\trans\ \langle \tvec{K} \rangle,
\label{J2}
\\
J_3 &=& \left(\tvec{\eta} \times (E \tvec{\eta}) \right)^\trans\ \langle \tvec{K} \rangle.
\label{J3}
\end{eqnarray}
See theorem 6 of~\cite{Man2} and~\cite{Lavoura:1994fv,Botella:1994cs}.
Here $\langle \tvec{K} \rangle$ denotes the $\tvec{K}$ vector formed by the
vacuum expectation values~$\langle \varphi_i \rangle$ of
the fields according to~\eqref{connect_4}.
Let us emphasize that in the case of more than one
CP invariance of the theory the statement
is that {\em at least one} CP invariance persists after EWSB if
$J_1=J_2=J_3=0$. 
In this case we call the Higgs potential
{\em spontaneously} CP conserving.

To summarize: the conditions~\eqref{I1}-\eqref{I4}, \eqref{J1}-\eqref{J3}
are relevant to study the {\em theoretical} GCP properties
of any given model:
i) if all $I's$ and $J's$ vanish, then there is at least one CP transformation
which is conserved at the Lagrangian level and also after EWSB.
Certainly, this CP symmetry would then be considered as {\em the}
CP symmetry of the theory.
This means that any CP violating experiment, such as a particle/antiparticle
decay asymmetry, will yield null results.
In the following we call this the {\em no CP violating} case -- but note the caveat below.
ii) If all $I's$ vanish and at least one $J$ does not, then there is spontaneous
CP violation.
iii) If at least one $I$ differs from zero, then the theory has explicit
CP violation.
We caution the reader regarding the following fact. A theory may have more than
one GCP transformation as a symmetry before EWSB. In this case the
vanishing of all $I's$ and $J's$ means that, out of these GCP symmetries,
at least one is preserved by the vacuum. As such, it will
be possible to define particles (scalars and pseudoscalars) which
are CP eigenstates, which affects the way they couple to the $Z$ boson,
for example. However this does {\em not} imply that {\em all} GCP symmetries
are likewise preserved. One such example is the model discussed
in~\cite{Maniatis:2007de} which has, at the Lagrangian level, four
GCP symmetries. Of these two are broken
by the vacuum-expectation value, while two other ones are
preserved.\\

The $J$ quantities involve the vacuum-expectation-values~(vevs)
of the fields.
If the theory has a neutral vacuum we may write without loss of generality
\begin{equation}
\label{vevs}
\begin{split}
\langle \varphi_1 \rangle &=
\begin{pmatrix}
0\\
v_1
\end{pmatrix}\\
\langle \varphi_2 \rangle &=
\begin{pmatrix}
0\\
v_2 e^{i\, \zeta}
\end{pmatrix},
\end{split}
\end{equation}
where $v_1$ and $v_2$ are real and the standard
Higgs vacuum-expectation value is $v_0= \sqrt{2(v_1^2 + v_2^2)} \approx 246$~GeV.
Inserting~\eqref{vevs} in \eqref{connect_4} we find
\begin{equation}
\langle \fvec{K} \rangle =
\begin{pmatrix}
\langle K_0 \rangle \\
\langle \tvec{K} \rangle
\end{pmatrix} =
\left(
\begin{array}{c}
v_0^2/2\\
2 v_1 v_2 \cos{\zeta}\\
2 v_1 v_2 \sin{\zeta}\\
v_1^2 - v_2^2
\end{array}
\right).
\end{equation}
The vacuum-expectation values are determined by
the stationarity conditions.
In the field language we need to set the gradient of the potential to zero.
In the bilinear formalism,
we must find the solutions of:
\begin{equation}
\left(
\fmat{E} - u \tilde{g}
\right)
\fvec{K}
=
-\frac{1}{2} \fvec{\xi}, \quad K_0^2 -\tvec{K}^2=0, \quad K_0>0 ;
\label{stat}
\end{equation}
see section~5 of~\cite{Man1}.
Here a Lagrange multiplier~$u$ is introduced in order to respect
the second constraint of~\eqref{stat}
and $\tilde{g}= \textrm{diag}(1,-1,-1,-1)$.
We have five equations
in five variables $K_0$, $K_1$, $K_2$, $K_3$, $u$, and one inequality.
Of course, one has to make sure to pick out from the
stationary points the absolute minimum of the potential as the vacuum
solution.\\

Note that here we focus on the CP properties of Higgs potentials.
For a study of their stability, electroweak symmetry breaking, and
of the solution of~\eqref{stat} corresponding to the global minimum
we refer to~\cite{Man1}. There, necessary and sufficient conditions
for physically required properties of the potential
are given in terms of the bilinear formalism.
We can add, however, that for all models discussed below - with both
exact and softly broken symmetries - we were able to find numerical
examples of viable minima. Thus, all classes of models below
are of physical interest.

\section{Symmetries of the THDM}
\label{secmodels}

The parameters in~\eqref{V_fields} may be reduced by imposing some symmetry
of the type $\varphi_i(x) \rightarrow S_{ij} \varphi_j(x)$,
or $\varphi_i(x) \rightarrow X_{ij} \varphi_j^\ast(x')$,
where $S$ and $X$ belong to $U(2)$.
The former are known as family symmetries,
the latter as generalized CP symmetries~\cite{Lee:1966ik,Ecker:1981wv,Ecker:1983hz,
Bernabeu:1986fc,Ecker:1987qp, Neufeld:1987wa}.
Recently it has been shown that applying the symmetries
with any possible choices for $S$ and $X$ leads only to six
classes of scalar potentials \cite{Ivanov:2007de,Ferreira:2009wh}.\\

Let us begin with the family symmetries.

\begin{itemize}
\item The $Z_2$ transformation~\cite{Glashow:1976nt,Paschos:1976ay} is defined as
\begin{equation}
\begin{pmatrix}
\varphi_1 \\ \varphi_2
\end{pmatrix}
\to
\begin{pmatrix}
1 & 0 \\ 0 & -1
\end{pmatrix}
\begin{pmatrix}
\varphi_1\\
\varphi_2
\end{pmatrix} .
\end{equation}
Using~\eqref{connect_4} we find the corresponding
transformation in $K$ space:
\begin{equation}
\label{Z2trans}
\begin{pmatrix} K_1 \\ K_2 \\ K_3 \end{pmatrix}
\to
\begin{pmatrix}
-1 & 0  & 0 \\
0 & -1  & 0 \\
0 & 0 & 1
\end{pmatrix}
\begin{pmatrix} K_1 \\ K_2 \\ K_3 \end{pmatrix}
\end{equation}
with $K_0$ unchanged.
That is, in $K$ space $Z_2$ is a rotation by
$\pi$ around the third axis. Using~\eqref{VK}
 and \eqref{eq-fourpar}
we see that $Z_2$ is a symmetry of the
potential $V$ if and only if
\begin{equation}
\tvec{\xi}=
\begin{pmatrix} 0\\ 0\\ \xi_3 \end{pmatrix}\;,
\qquad
\tvec{\eta}=
\begin{pmatrix} 0\\ 0\\ \eta_3 \end{pmatrix}\;,
\qquad
E=
\begin{pmatrix}
\eta_{11}& \eta_{12}& 0\\
\eta_{12}& \eta_{22}& 0\\
0 & 0 & \eta_{33} \\
\end{pmatrix}.
\end{equation}\\
Eventually, by a change of basis, we can achieve the
following form for the parameters of the Higgs potential
\begin{equation}
Z_2 : \quad
\tvec{\xi}=
\left(
\begin{array}{c}
0\\
0\\
\xi_3
\end{array}
\right),
\ \ \
\tvec{\eta}=
\left(
\begin{array}{c}
0\\
0\\
\eta_3
\end{array}
\right),
\ \ \
E =
\left(
\begin{array}{ccc}
\mu_1 &
0 &
0 \\
0 &
\mu_2 &
0 \\
0 &
0 &
\mu_3
\end{array}
\right).
\label{Z_2}
\end{equation}
A basis-invariant characterisation of this symmetry class
is that both $\tvec{\xi}$ and $\tvec{\eta}$ are
proportional to one and the same eigenvector of~$E$.

\item The Peccei--Quinn transformation $U(1)$
is defined as~\cite{Fayet:1974pd,Peccei:1977hh,Peccei:1977ur}
\begin{equation}
\label{eq24}
\begin{pmatrix}
\varphi_1 \\ \varphi_2
\end{pmatrix}
\to
\begin{pmatrix}
e^{- i \alpha} & 0 \\ 0 & e^{i \alpha}
\end{pmatrix}
\begin{pmatrix}
\varphi_1\\
\varphi_2
\end{pmatrix},
\qquad \text{with } 0 \le \alpha < \pi.
\end{equation}
In $K$ space
we get via~\eqref{connect_4}
\begin{equation}
\label{eq25}
\begin{pmatrix} K_1 \\ K_2 \\ K_3 \end{pmatrix}
\to
\begin{pmatrix}
\cos (2\alpha) & - \sin (2\alpha)  & 0 \\
\sin (2\alpha) &  \cos (2\alpha)  & 0 \\
0 & 0 & 1
\end{pmatrix}
\begin{pmatrix} K_1 \\ K_2 \\ K_3 \end{pmatrix} .
\end{equation}
That is, the $U(1)$ transformations
correspond to rotations around the third axis.

The potential $V$ is Peccei-Quinn symmetric if and
only if
there is a basis for which the parameters are
\begin{equation} \label{U1}
U(1): \quad \tvec{\xi}=
\begin{pmatrix} 0\\ 0\\ \xi_3 \end{pmatrix},
\quad
\tvec{\eta}=
\begin{pmatrix} 0\\ 0\\ \eta_3 \end{pmatrix},
\quad
E=
\begin{pmatrix}
\mu_{1}& 0 & 0\\
0 & \mu_{1}& 0\\
0 & 0 & \mu_{3} \\
\end{pmatrix},
\end{equation}
as we easily see, combining~\eqref{VK}, \eqref{eq-fourpar}, and \eqref{eq25}.
Here a basis independent formulation is that 
two eigenvalues of~$E$ must be degenerate and both vectors
$\tvec{\xi}$ and $\tvec{\eta}$ must be proportional to
the eigenvector corresponding to the remaining third
eigenvalue of~$E$.
\item The $U(2)$ transformations in field space are given by
\begin{equation}
\varphi_i(x) \rightarrow S_{ij} \varphi_j(x) , \qquad i=1,2 ,
\end{equation}
with $S=(S_{ij}) \in U(2)$.
In $K$ space this corresponds to
\begin{equation}
\tvec{K} \to R \; \tvec{K}, \qquad \text{with } R \in SO(3) .
\end{equation}
The Higgs potential, invariant under all such transformations,
has parameters
\begin{equation}
U(2) : \quad
\tvec{\xi}=
\left(
\begin{array}{c}
0\\
0\\
0
\end{array}
\right),
\ \ \
\tvec{\eta}=
\left(
\begin{array}{c}
0\\
0\\
0
\end{array}
\right),
\ \ \
E =
\left(
\begin{array}{ccc}
\mu_1 &
0 &
0 \\
0 &
\mu_1 &
0 \\
0 &
0 &
\mu_1
\end{array}
\right).
\label{U2}
\end{equation}
Formulated in a basis-independent way we have
$\tvec{\xi}=0$ and $\tvec{\eta}=0$
and~$E$ proportional to the unit matrix.
\end{itemize}

Now we come to the models which have
generalized CP symmetries of the different types~\eqref{GCPclass}.

\begin{itemize}
\item The CP1 transformations with standard form~$\theta=0$ in~\eqref{eq32}
correspond in $K$ space to reflections on
planes in addition to the parity transformation of the argument.
By a suitable basis change, any CP1 reflection
can be rotated to occur on the 1--3 plane,
as given in~\eqref{standCPK}.
The parameters of a Higgs potential, invariant under a CP1 transformation,
can, with a suitable basis change, always be brought to the form
\begin{equation}
\text{CP1}: \quad
\tvec{\xi}=
\left(
\begin{array}{c}
\xi_1\\
0\\
\xi_3
\end{array}
\right),
\ \ \
\tvec{\eta}=
\left(
\begin{array}{c}
\eta_1\\
0\\
\eta_3
\end{array}
\right),
\ \ \
E =
\left(
\begin{array}{ccc}
\mu_1 &
0 &
0 \\
0 &
\mu_2 &
0 \\
0  &
0 &
\mu_3
\end{array}
\right).
\label{CP1}
\end{equation}
Formulated in a basis-independent way this
means that $\tvec{\xi}$ and $\tvec{\eta}$
must be orthogonal to one and the same
eigenvector of~$E$.

\item Generalized CP transformations of type CP2 correspond in
$K$ space to the point reflection at the origin in addition
to the parity transformation of the argument,
\begin{equation}
\tvec{K}(x) \to -\mathbbm{1}_3 \tvec{K}(x') .
\end{equation}
Therefore, the Higgs potential parameters with this
symmetry are, in a basis where $E$ is diagonal~\cite{Davidson:2005cw,Man2},
\begin{equation}
\text{CP2}: \quad
\tvec{\xi}=
\left(
\begin{array}{c}
0\\
0\\
0
\end{array}
\right),
\ \ \
\tvec{\eta}=
\left(
\begin{array}{c}
0\\
0\\
0
\end{array}
\right),
\ \ \
E =
\left(
\begin{array}{ccc}
\mu_1 &
0 &
0 \\
0 &
\mu_2 &
0 \\
0  &
0 &
\mu_3
\end{array}
\right).
\label{CP2}
\end{equation}
The basis-independent requirement is here $\tvec{\xi}=0$ and
$\tvec{\eta}=0$.

\item Finally, the generalized CP transformations of type CP3 correspond
in $K$ space to improper rotations $\bar{R}$ in~\eqref{GCPK} with
$0 < \theta < \pi/2$ in addition to $x \to x'$.
The Higgs potential, invariant under CP3, has the parameters
\begin{equation}
\text{CP3}: \quad
\tvec{\xi}=
\left(
\begin{array}{c}
0\\
0\\
0
\end{array}
\right),
\ \ \
\tvec{\eta}=
\left(
\begin{array}{c}
0\\
0\\
0
\end{array}
\right),
\ \ \
E =
\left(
\begin{array}{ccc}
\mu_1 &
0 &
0 \\
0 &
\mu_2 &
0 \\
0  &
0 &
\mu_1
\end{array}
\right).
\label{CP3}
\end{equation}
The basis-independent formulation reads here
$\tvec{\xi}=0$ and $\tvec{\eta}=0$, and
two eigenvalues of $E$ degenerate.
\end{itemize}

The impact of the six classes on the parameters in terms
of the fields is
given in Table~\ref{master1}, for a specific basis choice.
%
\begin{table*}[ht!]
\caption{\em Impact of the symmetries on the coefficients
of the Higgs potential in a specific basis.
See Ref.~\cite{Ferreira:2009wh} for more details.}
\begin{tabular}{ccccccccccc}
\hline
\hline
symmetry & $m_{11}^2$ & $m_{22}^2$ & $m_{12}^2$ &
$\lambda_1$ & $\lambda_2$ & $\lambda_3$ & $\lambda_4$ &
$\lambda_5$ & $\lambda_6$ & $\lambda_7$ \\
\hline
$Z_2$ &   &   & 0 &
   &  &  &  &
real & 0 & 0 \\
$U(1)$ &  &  & 0 &
 &  & &  &
0 & 0 & 0 \\
$U(2)$ &  & $ m_{11}^2$ & 0 &
   & $\lambda_1$ &  & $\lambda_1 - \lambda_3$ &
0 & 0 & 0 \\
\hline
CP1 &  &  & real &
 & &  &  &
real & real & $\lambda_6$ \\
CP2 &  & $m_{11}^2$ & 0 &
  & $\lambda_1$ &  &  &
real & 0 & 0 \\
CP3 &  & $m_{11}^2$ & 0 &
   & $\lambda_1$ &  &  &
$\lambda_1 - \lambda_3 - \lambda_4$ (real) & 0 & 0 \\
\hline
\hline
\end{tabular}
\label{master1}
\end{table*}
%
The first three may be obtained from family symmetries alone.
The next three may be obtained from GCP symmetries alone.
The CP2 entry is presented in a very specific basis
mentioned by Davidson and Haber \cite{Davidson:2005cw}.
For $Z_2$,
$\lambda_5$ may be made real without loss of generality.
In appendix~\ref{multi} we discuss
cases where two symmetries are imposed simultaneously.
\\

The models $Z_2$, $U(1)$, $U(2)$, CP1, CP2, and CP3
discussed so far obey exact symmetries.
We shall also consider the case that these
symmetries are broken by
including all possible Higgs-potential terms of dimension two.
This is known as soft breaking of the symmetry.
Soft-breaking terms arise for instance if we consider a
THDM as an effective model. For example, in the minimal supersymmetric
extension of the Standard Model (MSSM), supersymmetry is
broken by imposing soft-breaking terms.
In this way mass degeneracy of the particles and their 
superpartners is
avoided. Another example is the 
next-to-minimal supersymmetric Standard Model~(NMSSM), wherein the $\mu$
term is generated by the vacuum expectation value of a singlet field.
For a review of the NMSSM see, for instance, \cite{Maniatis:2009re}.
If the first symmetry breaking occurs at an energy scale~$v_s$
considerably larger than
the electroweak scale~$v_0$, we obtain below the scale~$v_s$
an effective THDM with 
soft-breaking terms~\cite{Ma:2010ya}.
This type of NMSSM model is in fact a particular example
of theories containing two doublets and additional singlet
fields where the latter acquire vevs and in this way produce a low-energy
effective theory analogous to the softly-broken THDM's discussed here.
In general, soft breaking of a symmetry is
employed in order to avoid unwanted massless Goldstone
modes in case the symmetry is continuous or in order
to avoid the domain-wall problem in case of a discrete symmetry.
A further motivation to consider soft breaking is as follows.
Theories with a
softly broken symmetry preserve under at least one-loop
renormalization whatever relations between the
quartic parameters were imposed by the exact symmetry.

In theories with general soft breaking,
$m_{11}^2$,
$m_{22}^2$,
$\re(m_{12}^2)$, and $\im(m_{12}^2)$ in the potential~\eqref{V_fields}
do not vanish and bear no relation to each other.
According to~\eqref{connect_1}, this corresponds
in $K$ space to a non-vanishing vector
\begin{equation}
\tvec{\xi}=
\left(
\begin{array}{c}
\xi_1\\
\xi_2\\
\xi_3
\end{array}
\right).
\label{soft_xi}
\end{equation}
Thus, in the following we shall
study THDM's with softly broken $Z_2,...$,~CP3 symmetries,
that is, THDM's where the parameters $\tvec{\eta}$ and $E$
are for the various classes as
in~\eqref{Z_2}, \eqref{U1}, \eqref{U2}, \eqref{CP1}, \eqref{CP2}, \eqref{CP3},
respectively,
but $\tvec{\xi}$ is left arbitrary.

\section{CP properties}

The main reason why T.D. Lee introduced the THDM in 1973~\cite{Lee:1973iz}
was precisely because one could have spontaneous CP breaking in
the model. In fact it is well known that spontaneous CP breaking
occurs in the $Z_2$ model with soft breaking~\cite{Branco:1985aq} or in
the CP1 model, but that it cannot occur in the unbroken
$Z_2$ and $U(1)$ cases;
see, for instance,~\cite{Velhinho:1994vh,Ferreira:2004yd}.

In this section we will show that the powerful bilinear formalism,
coupled with the equally powerful basis-invariant methods, allows
us to investigate CP breaking, explicit or spontaneous,
for all symmetry-constrained THDM Higgs potentials discussed in the
last section.
As we will see this is possible
for the most part, even without computing the vacuum-expectation values.
We must however emphasize that in the following discussion of the various
models we always assume that the parameters~\eqref{connect_1}--\eqref{connect_3}
of the potential are chosen such that we have
stability and a neutral vacuum different from~$\langle \tilde{\tvec{K}} \rangle=0$.

We note that we have obtained
numerical examples with these properties for each case discussed below.
Since the conditions for the CP symmetries 
\eqref{I1}-\eqref{I4},
\eqref{J1}-\eqref{J3}
are 
basis independent, we can apply them to models
given in any basis.
We use this freedom to consider the models
in convenient bases -- for instance in bases where 
the parameter matrix~$E$ is diagonal.
Since the conditions are not affected by 
a change of basis the results 
hold for the models basis independently.
It is also straightforward to formulate the results written down below
in the basis where $E$ is diagonal in an explicitly basis-invariant way:
clearly, the eigenvalues of $E$ are basis-invariant quantities.
Furthermore, suppose that we use the components
$\xi_1$, $\xi_2$, $\xi_3$ of the vector $\tvec{\xi}$ in
the $E$-diagonal basis. In a basis-independent formulation these 
components of $\tvec{\xi}$ are the projections of $\tvec{\xi}$ onto the normalized
eigenvectors of $E$. 
The same holds for these components of the other vectors, $\tvec{\eta}$
and $\langle \tvec{K} \rangle$.
In the following explicitly basis-independent formulations are always
easily obtainable in this way.

\subsection*{The $U(2)$ model}

\subsubsection*{Explicit CP breaking}

In the unbroken $U(2)$ model we have $\tvec{\xi}=\tvec{\eta} = 0$,
thus from~\eqref{I1}--\eqref{I4} follows
$I_1 = I_2 = I_3 = I_4 = 0$, that is,
we have explicit CP conservation.
Even in the soft breaking case, with~$\tvec{\xi}$ arbitrary, we find
from $\tvec{\eta}=0$ with~\eqref{I1}--\eqref{I4} immediately that $I_1 = I_2 = I_4 = 0$.
Moreover, since
the matrix $E$ is proportional to the identity, the vector $E \tvec{\xi}$
is parallel to $\tvec{\xi}$ and thus, from \eqref{I3}, we also obtain
$I_3 = 0$, with or without soft breaking terms.
We find that there is at least one explicit CP symmetry in the $U(2)$ model,
even in the presence of the soft-breaking terms.

\subsubsection*{Spontaneous CP breaking}

Since $\tvec{\eta} = 0$ we have $J_1 = J_3 = 0$ and since $\tvec{\xi}$
and $E \tvec{\xi}$ vanish for the exact symmetry and are parallel in the soft-breaking case,
$J_2$ is also zero, as can be seen from~\eqref{J1}
to \eqref{J3}.
Hence, at least one CP symmetry is conserved by
the vacuum, even if we take the
most general soft breaking terms into account.

\subsection*{The CP3 model}

\subsubsection*{Explicit CP breaking}

The main difference with
regard to the $U(2)$ model is that now~$E$ is not proportional to the identity -- only
two of the eigenvalues are degenerate.
The unbroken CP3 model implies that $\tvec{\eta} = \tvec{\xi} = 0$, 
so we get here $I_1 = I_2 = I_3 = I_4 = 0$.
Let us turn to the softly broken case with a non-vanishing vector~$\tvec{\xi}$.
Since $\tvec{\eta}=0$, we find immediately $I_1=I_2=I_4=0$.
The invariant $I_3$ reads in any basis with a diagonal matrix~$E$
\begin{equation}
I_3 = \xi_1\, \xi_2\, \xi_3\,
(\mu_1 - \mu_2)
(\mu_2 - \mu_3)
(\mu_3 - \mu_1) .
\label{I3_useful}
\end{equation}
From the degeneracy of two eigenvalues of~$E$ follows that
$I_3=0$. We see this also directly from~\eqref{I3}.
Because~$E$ has two degenerate eigenvalues
the vectors~$\tvec{\xi}$, $E \tvec{\xi}$, $E^2 \tvec{\xi}$ lie in
one plane, that is, the triple product in $I_3$ vanishes.
Thus, the CP3 model is explicitly CP conserving, even in the
softly broken case~\footnote{In this paper we are concerned only with the
scalar potential. However, should one include the fermionic sector, it has been proven
that the softly broken CP3 model has strange CP violating properties~\cite{Ferreira:2010bm}.}.

\subsubsection*{Spontaneous CP breaking}
In the unbroken model we have $\tvec{\eta} = \tvec{\xi} = 0$ and get
immediately from~\eqref{J1} to \eqref{J3}
that all $J's$ are zero. In the softly broken case, with $\tvec{\xi}$ arbitrary,
we still have $\tvec{\eta}=0$, that
is, $J_1$ and $J_3$ are zero. Since $\mu_1 = \mu_3$, we may perform
a basis transformation such that $\xi_3 = 0$, leaving
$\tvec{\eta}$ and $E$ unchanged. With this simplification, $J_2$ is given by
\begin{equation}
\label{SCP3}
J_2 = \xi_1 \xi_2 (\mu_2 - \mu_1) \langle K_3 \rangle.
\end{equation}
In order to derive further conclusions we look at the
stationarity conditions~\eqref{stat}. We get explicitly
\begin{equation}
\label{solCP3}
\begin{pmatrix}
\langle K_0 \rangle (\eta_{00} -u)\\
\langle K_1 \rangle (\mu_1 + u)\\
\langle K_2 \rangle (\mu_2 +u )\\
\langle K_3 \rangle (\mu_1 +u)
\end{pmatrix}
=
- \frac{1}{2}
\begin{pmatrix}
\xi_0\\
\xi_1\\
\xi_2\\
0\\
\end{pmatrix} .
\end{equation}
For the case $\mu_1+u \neq 0$ there is
only a solution for~$\langle K_3 \rangle = 0$.
On the other hand, for~$\mu_1+u = 0$,
we get~$\xi_1=0$. Thus we see from~\eqref{SCP3} that in
either case the condition $J_2=0$ is fulfilled. Hence,
for the unbroken and softly broken CP3 model, there is at least one CP symmetry respected by the vacuum.

\subsection*{The $U(1)$ model}

\subsubsection*{Explicit CP breaking}

Consider first the exactly symmetric model with parameters~\eqref{U1}.
The vectors $\tvec{\eta}$ and $\tvec{\xi}$ are not zero, but they
are parallel to each other, as are $E \tvec{\eta}$ and $E \tvec{\xi}$. This means
that all the $I's$ are zero, as is easy to see
from~\eqref{I1}--\eqref{I4}. In the softly broken case,
since we have two degenerate eigenvalues
in the matrix~$E$, we find that
still all $I's$ vanish; see~\eqref{I3_useful}.
Therefore, we have explicit CP conservation with the $U(1)$ scalar
potential, even if we take soft breaking terms into account.

\subsubsection*{Spontaneous CP breaking}
For the unbroken model, since $\tvec{\eta}$, $\tvec{\xi}$,
$E \tvec{\xi}$, $E \tvec{\eta}$ are parallel, all $J's$ are zero.
For the softly broken $U(1)$ case we can,
without loss of generality, go into a basis with $\xi_2=0$.
We get then for the $J's$
from~\eqref{J1} to \eqref{J3}
\begin{equation}
\label{SCPU1}
J_1 = - \xi_1 \eta_3 \langle K_2 \rangle, \quad
J_2 = \xi_1 \xi_3 (\mu_1 - \mu_3) \langle K_2 \rangle, \quad
J_3 = 0.
\end{equation}
Therefore, we need to look at the vacuum properties before
we can reach further conclusions.
Employing again the stationarity conditions~\eqref{stat},
we get
\begin{equation}
\label{solU1}
\begin{pmatrix}
\langle K_0 \rangle (\eta_{00} -u) + \langle K_3 \rangle \eta_3\\
\langle K_1 \rangle (\mu_1 + u)\\
\langle K_2 \rangle (\mu_1 +u )\\
\langle K_3 \rangle (\mu_3 +u) + \langle K_0 \rangle \eta_3
\end{pmatrix}
=
- \frac{1}{2}
\begin{pmatrix}
\xi_0\\
\xi_1\\
0\\
\xi_3\\
\end{pmatrix} .
\end{equation}
This means that we get for~$\mu_1+u \neq 0$ directly
$\langle K_2 \rangle = 0$ and for $\mu_1+u = 0$ we have
$\xi_1=0$. We see from~\eqref{SCPU1}, that
in either case
all $J's$ are zero and we have at least one CP symmetry
which is preserved by the vacuum.

\subsection*{The CP2 model}

\subsubsection*{Explicit CP breaking}
Here, for the unbroken case, $\tvec{\eta} = \tvec{\xi} = 0$ implies
again that all the $I$'s are zero, that is, we have explicit CP
conservation, as trivially expected for a CP invariant model.
In the softly broken case, with an arbitrary vector~$\tvec{\xi}$,
only $I_3$ as given in~\eqref{I3_useful} may be non-zero.
In the case of the softly broken CP2 model with non-degenerate eigenvalues
of $E$, we thus in general do not have explicit CP conservation.

However, in the softly-broken CP2 model it is still possible
to choose the soft-breaking parameters such that
CP is explicitly conserved.
For instance, in the case that at least one of the components
of~$\tvec{\xi}$ is zero (in a basis where $E$ is diagonal) we have explicit CP invariance.
It is interesting to consider a few particular cases:
if $\xi_1 = \xi_2 = 0$ (and $\xi_3 \neq 0$),
corresponding to $\textrm{Re} (m_{12}^2) = \textrm{Im} (m_{12}^2) =0$,
then we see from a comparison of~\eqref{CP2} with \eqref{Z_2}
or from Table~\ref{master1} that the softly-broken CP2
model has a remaining $Z_2$ symmetry
($\varphi_1 \rightarrow \varphi_1; \varphi_2 \rightarrow - \varphi_2$).
Similarly,
if $\xi_2 = \xi_3 = 0$ (and $\xi_1 \neq 0$),
corresponding to $\textrm{Im} (m_{12}^2) = m_{22}^2 - m_{11}^2 = 0$,
then the softly-broken CP2 model has a remaining $\Pi_2$ symmetry
($\varphi_1 \leftrightarrow \varphi_2$, which is simply $Z_2$ in a different basis; see appendix~\ref{multi}).
Furthermore, if we choose
$\xi_2 = 0$ ($\textrm{Im} (m_{12}^2) = 0$) we see from~\eqref{I3_useful}
that the model will have an unbroken CP symmetry.
Note that the cases of degenerate eigenvalues in the matrix~$E$ correspond
actually to the CP3 model (maybe in a different basis) or to the $U(2)$ model.
All of these symmetries - $Z_2$, $\Pi_2$, and CP1 - if exact, imply
explicit CP conservation.

\subsubsection*{Spontaneous CP breaking}

For the unbroken case, $\tvec{\eta} = \tvec{\xi} = 0$,
we see from~\eqref{J1} to \eqref{J3} that all $J$'s are 
trivially zero, that is we have
spontaneous CP conservation.
For the softly broken model, $\tvec{\xi}\neq 0$, consider the case of
explicit CP conservation, that is, the case with all $I's$,
in particular $I_3$ in \eqref{I3_useful},
vanishing. We have then $J_1=J_3=0$, and $J_2$ given by
\begin{equation}
J_2 =
\xi_2\, \xi_3\, (\mu_3 - \mu_2) \langle K_1 \rangle
+
\xi_1\, \xi_3\, (\mu_1 - \mu_3) \langle K_2 \rangle
+
\xi_1\, \xi_2\, (\mu_2 - \mu_1) \langle K_3 \rangle.
\label{J2_useful}
\end{equation}
In order to reach further conclusions about the spontaneous
breaking behavior we consider the stationarity conditions~\eqref{stat},
\begin{equation}
\label{solCP2}
\begin{pmatrix}
\langle K_0 \rangle (\eta_{00} -u)\\
\langle K_1 \rangle (\mu_1 + u)\\
\langle K_2 \rangle (\mu_2 +u )\\
\langle K_3 \rangle (\mu_3 +u)
\end{pmatrix}
=
- \frac{1}{2}
\begin{pmatrix}
\xi_0\\
\xi_1\\
\xi_2\\
\xi_3\\
\end{pmatrix} .
\end{equation}
We see that for a general solution of the stationarity conditions we do not
necessarily find $J_2 =0$ in the soft breaking case, that
is, we have in general spontaneous CP violation.

Remember that it is sufficient to have one of the components
of~$\tvec{\xi}$ equal to zero in order to have explicit CP conservation
in the CP2 model.
For instance, if $\xi_2 = 0$ and $u\neq -\mu_2$, then
$\langle K_2 \rangle = 0$ and $J_2 = 0$. But the solution \mbox{$u =-\mu_2$}
and $\langle K_2 \rangle \neq 0$
is also a possibility, and this corresponds to a vacuum with spontaneous CP
violation.

To summarize: a model with exact CP2 symmetry is explicitly and spontaneously
CP conserving, that is, at least one CP symmetry is conserved explicitly
and by the vacuum. This was
already proven in~\cite{Man2}. For the softly broken case
the parameters can be chosen such as to have explicit
CP conservation. For the case of explicit CP conservation we may
have spontaneous CP violation.

\subsection*{The $Z_2$ model}

\subsubsection*{Explicit CP breaking}

This case is quite similar to that of the $U(1)$ model: for the unbroken model
 the vectors $\tvec{\eta}$, $\tvec{\xi}$,
$E \tvec{\eta}$ and $E \tvec{\xi}$ are all parallel to each other and, therefore,
the $I$ invariants are zero. For the softly broken case we still have
$I_2 = I_4 = 0$, $I_3$ is given in~\eqref{I3_useful}, and
\begin{equation}
I_1 = \xi_1\, \xi_2\, \eta_3\, (\mu_1 - \mu_2).
\label{I1_useful}
\end{equation}
Thus, in general the softly broken $Z_2$ model is explicitly CP violating.
But, obviously, in particular cases it is possible to have explicit CP conservation.
For instance, for the case that one of the parameters~$\xi_1$ or $\xi_2$ is equal
to zero, or that the two eigenvalues $\mu_1$ and $\mu_2$ of the matrix~$E$ are degenerate,
we find from~\eqref{I3_useful} and~\eqref{I1_useful} that all $I's$ vanish.
But one must study whether that reproduces one of the other five
symmetry-constrained models. For example, starting from the softly-broken $Z_2$ model and imposing
$\mu_1=\mu_2$ we recover the explicitly CP conserving softly-broken $U(1)$ model.
Likewise, for $\eta_3=0$ we end up with the softly broken CP2 model.
Thus, in studying the $Z_2$ model with soft breaking
and explicit CP conservation we may
focus on the cases where either $\xi_1$ or $\xi_2$ is zero.

\subsubsection*{Spontaneous CP breaking}

For the unbroken $Z_2$ model, the vectors~$\tvec{\xi}$, $\tvec{\eta}$,
$E\tvec{\xi}$, $E\tvec{\eta}$
entering into the computation
of the $J$ invariants in \eqref{J1}--\eqref{J3} are parallel. Thus, all $J's$ vanish.
The exact $Z_2$ symmetry implies spontaneous CP conservation,
as is well known; see for example~\cite{Branco:1985aq,Velhinho:1994vh,Ferreira:2004yd}.

Concerning the softly broken $Z_2$ model suppose that we have explicit CP conservation, that
is, the parameters are such that besides $I_2$ and $I_4$, also $I_3$~\eqref{I3_useful}
and $I_1$~\eqref{I1_useful}
vanish. As already discussed, we can disregard the cases
$\mu_1=\mu_2$ and $\eta_3=0$  because
they correspond to the $U(1)$ and CP2 models.
Therefore, we must have~$\xi_1=0$ or
$\xi_2=0$.
We find then $J_3 = 0$, $J_2$ is
given by~\eqref{J2_useful} and
\begin{equation}
J_1 = \eta_3 (\xi_2 \langle K_1 \rangle - \xi_1 \langle K_2 \rangle).
\label{J2Z2}
\end{equation}
The vacuum solution is determined by
\begin{equation}
\label{solZ2}
\begin{pmatrix}
\langle K_0 \rangle (\eta_{00} -u) + \langle K_3 \rangle \eta_3\\
\langle K_1 \rangle (\mu_1 + u)\\
\langle K_2 \rangle (\mu_2 +u )\\
\langle K_3 \rangle (\mu_3 +u) + \langle K_0 \rangle \eta_3
\end{pmatrix}
=
- \frac{1}{2}
\begin{pmatrix}
\xi_0\\
\xi_1\\
\xi_2\\
\xi_3\\
\end{pmatrix} .
\end{equation}
We see that even in the case where either $\xi_1$ or $\xi_2$ is equal to zero,
we still have that $J_1$ and $J_2$ are not necessarily zero. For
instance, for $\xi_2 = 0$, $\xi_1 \neq 0$, and $u = -\mu_2$, we can obtain
$\langle K_2 \rangle \neq 0$, that is, $J_1 \neq 0$.
Thus, spontaneous CP violation is possible in softly broken $Z_2$ models,
even though it cannot occur in the case of the exact symmetry.
This result is well-known in fact and was shown in~\cite{Branco:1985aq},
where a $\Pi_2$ symmetry was considered,
which is just the $Z_2$ model in a different basis.

\subsection*{The CP1 model}

\subsubsection*{Explicit CP breaking}

The unbroken CP1 model has vanishing $I's$. It
is, by construction, explicitly CP conserving.
The softly broken CP1 model arises from the unbroken case by
having~$\xi_2 \neq 0$. We can in addition assume that
both $\eta_1 \neq 0$ and $\eta_3 \neq 0$, since
otherwise we get the $Z_2$ case (or the $\Pi_2$ case, that is, $Z_2$ in a different basis).
The explicit CP conditions~\eqref{I1}-\eqref{I4} read then~$I_4=0$,
\begin{equation}
\begin{split}
I_1 &=\xi_2 \big[
  \xi_1 \eta_3 (\mu_1 -\mu_2)
+ \xi_3 \eta_1 (\mu_2 - \mu_3) \big]=0,\\
I_2 &= \xi_2 \eta_1 \eta_3 (\mu_1 - \mu_3)=0 ,\\
I_3 &= \xi_1 \xi_2 \xi_3 (\mu_1 - \mu_2)
(\mu_2 - \mu_3) (\mu_3 - \mu_1) =0.
\end{split}
\end{equation}
Thus, in general, we will have explicit CP violation in the softly broken CP1 model.

\subsubsection*{Spontaneous CP breaking}

Suppose that CP is explicitly conserved.
According to the discussion above this implies that we have the unbroken
CP1 model where~$\xi_2=0$.
The $J$ invariants have then the form
\begin{equation}
\begin{split}
J_1 & = (\xi_3 \eta_1 - \xi_1 \eta_3) \; \langle K_2 \rangle ,\\
J_2 & = \xi_1\, \xi_3\, (\mu_1 - \mu_3) \; \langle K_2 \rangle  , \\
J_3 & = \eta_1\, \eta_3\, (\mu_1 - \mu_3)\; \langle K_2 \rangle  .
\end{split}
\end{equation}
The stationarity conditions~\eqref{stat} are
\begin{equation}
\label{solCP1}
\begin{pmatrix}
\langle K_0 \rangle (\eta_{00} -u) + \langle K_1 \rangle \eta_1 + \langle K_3 \rangle \eta_3\\
\langle K_1 \rangle (\mu_1 + u) + \langle K_0 \rangle \eta_1\\
\langle K_2 \rangle (\mu_2 +u )\\
\langle K_3 \rangle (\mu_3 +u) + \langle K_0 \rangle \eta_3
\end{pmatrix}
=
- \frac{1}{2}
\begin{pmatrix}
\xi_0\\
\xi_1\\
0\\
\xi_3\\
\end{pmatrix} .
\end{equation}
We see that a valid vacuum may occur with
$\langle K_2 \rangle \neq 0$ with a Lagrange multiplier $u=-\mu_2$.
Then, the three $J's$ do in general not vanish.
Thus, the CP1 model may have spontaneous CP violation.
This result was obtained by T.~D.~Lee
a long time ago~\cite{Lee:1973iz}.

\section*{Summary}

We present an overview of the CP properties
of the six symmetry-constrained THDM's,
with and without soft symmetry breaking,
in Table~\ref{master2}, which is the main
result of this article.
%
\begin{table*}[ht!]
\caption{\em CP properties of the six symmetry-constrained THDM's,
with and without soft-symmetry breaking.
``Yes'' means that it is possible to choose the parameters of
the potential such as to enable that particular form of CP violation.}
\begin{center}
\begin{tabular}{|c|c|c|c|c|}
\hline
\hline
  & \multicolumn{2}{|c|}{exact} & \multicolumn{2}{|c|}{softly-broken} \\
\cline{2-5}
 symmetry & explicit & spontaneous & explicit & spontaneous  \\
          & CPV & CPV & CPV & CPV  \\
\hline
$Z_2$ &  -- & -- & Yes & Yes \\
$U(1)$ & -- & -- & -- & -- \\
$U(2)$ & -- & -- & -- & -- \\
\hline
CP1 & -- & Yes & Yes & Yes \\
CP2 & -- & -- & Yes & Yes \\
CP3 & -- & -- & -- & -- \\
\hline
\hline
\end{tabular}
\end{center}
\label{master2}
\end{table*}
%
Table~\ref{master2} displays the whole
gamut of possibilities of CP breaking in any of the six possible classes of
symmetry-constrained THDM's. As we have seen, it was possible to obtain
these results without the need
to determine explicit expressions for the vacuum-expectation values.
Regarding the last two
columns, please be aware that any model which can display explicit CP breaking
therein listed can {\em also}, through a judicious choice of the parameters,
explicitly preserve CP -- and, in that case, develop vacua which spontaneously
break CP, which is indicated in the last column.

\section{Conclusions}

In this work we have performed a thorough study of the CP breaking
properties, both explicit and spontaneous, of all possible THDM scalar potentials
corresponding to any of the six symmetry classes.
We also considered the softly broken
models. Our study was performed
in a very efficient manner using basis-invariant quantities written
in the bilinear formalism, which simplified the analysis immensely. The
results are presented in Table~\ref{master2}. This includes some
results scattered through the literature, sometimes only implicitly,
as well as our new results.
Table~\ref{master2} allows us to draw some general conclusions:
\begin{itemize}
\item {\em Any} exact symmetry prevents explicit CP breaking.
\item All the models which possess a {\em continuous} symmetry -- $U(1)$, $U(2)$,
and CP3 -- cannot have explicit or spontaneous CP breaking, even with the most
general soft  breaking  terms.
\item All models with {\em discrete} symmetries -- $Z_2$, CP1, and CP2 --
may have explicit CP violation, but they require
soft symmetry breaking terms to make it happen.
\item In models with exact symmetries~$Z_2$ or CP2 there
is no spontaneous CP violation.
But we remind the reader that this only means that at least
one CP symmetry is preserved by the vacuum.
The unbroken CP2 model has four GCP symmetries where two
are broken and two are unbroken if the theory has the
correct EWSB; see~\cite{Man2}.
\item In the models with softly broken {\em discrete} symmetries,
$Z_2$ or CP2,
we may choose the parameters in a way to achieve explicit CP conservation.
Then these models may have spontaneous CP violation.
\item In models with exact CP1 symmetry, spontaneous
CP violation is possible as has been known for a long time~\cite{Lee:1973iz}.
Inclusion of soft-breaking terms either violates the CP symmetry
explicitly
or leads back to the exactly symmetric case.
\end{itemize}
We remind the reader that our analysis concerned only the scalar sector, nothing
was said about CP violation in the Yukawa sector. There one can
have explicit CP violation even if the scalar potential is explicitly CP conserving (the SM is a prime example of this). Also, a
Lagrangian where the Higgs potential preserves CP explicitly and spontaneously
may well break CP spontaneously in the fermion sector. 
An example of this unusual phenomenon
was found in~\cite{Ferreira:2010bm}.

\acknowledgments{
The work of P.M.F. is supported in part by the Portuguese
\textit{Funda\c{c}\~{a}o para a Ci\^{e}ncia e a Tecnologia} (FCT)
under contract PTDC/FIS/70156/2006.
The work of J.P.S. is funded by FCT through the
projects CERN/FP/109305/2009 and  U777-Plurianual,
and by the EU RTN project Marie Curie: MRTN-CT-2006-035505.
The work of M.M. was funded by Deutsche Forschungsgemeinschaft, project
number NA296/5-1.
P.F. thanks the gracious hospitality
of the Institut f\"ur Theoretische Physik at Heidelberg during part of this
work. M.M. thanks the Centro de F\'{\i}sica Te\'orica e Computacional at
Universidade de Lisboa for its hospitality during his visit to Lisbon.}

\appendix

\section{Multiple Higgs-family symmetries}
\label{multi}

In~\cite{Ferreira:2009wh} it was shown that the simultaneous
imposition of two Higgs-family symmetries corresponds
in a certain basis to another Higgs-family symmetry in some cases.
Here we want to show these results and similar ones in terms of
the bilinear formalism~\cite{Man1,Man2}.
But let us first recall the $\Pi_2$ transformation
\begin{equation}
\varphi_1 \leftrightarrow \varphi_2
\end{equation}
which corresponds in $K$ space to a rotation by~$\pi$
around the first axis
\begin{equation}
\begin{pmatrix} K_1 \\ K_2 \\ K_3 \end{pmatrix}
\to
\begin{pmatrix}
1 & 0  & 0 \\
0 & -1  & 0 \\
0 & 0 & -1
\end{pmatrix}
\begin{pmatrix} K_1 \\ K_2 \\ K_3 \end{pmatrix} .
\end{equation}
Clearly, by a basis change $\Pi_2$ is equivalent
to~$Z_2$,~\eqref{Z2trans}.

Now we discuss some cases of multiple symmetries:

\begin{itemize}
\item Simultaneous $U(1)$ and CP3 symmetries.
From~\eqref{U1} and \eqref{CP3} we see, that in the
basis chosen there
both symmetries combined give the
$U(2)$ model~\eqref{U2}. Thus, we confirm that
in a certain basis $U(1) \oplus \text{CP3} = U(2)$.

\item Next we introduce the $\Pi_2$ symmetric model,
which, by a basis change, is equivalent to $Z_2$:
\begin{equation}
\label{Pi2}
\Pi_2: \quad
\tvec{\xi}=
\left(
\begin{array}{c}
\xi_1\\
0\\
0
\end{array}
\right),
\ \ \
\tvec{\eta}=
\left(
\begin{array}{c}
\eta_1\\
0\\
0
\end{array}
\right),
\ \ \
E =
\left(
\begin{array}{ccc}
\mu_1 &
0 &
0 \\
0 &
\mu_2 &
0 \\
0 &
0 &
\mu_3
\end{array}
\right).
\end{equation}
Applying simultaneously $U(1)$ and $\Pi_2$ we get
\begin{equation}
\tvec{\xi}=
\left(
\begin{array}{c}
0\\
0\\
0
\end{array}
\right),
\ \ \
\tvec{\eta}=
\left(
\begin{array}{c}
0\\
0\\
0
\end{array}
\right),
\ \ \
E =
\left(
\begin{array}{ccc}
\mu_1 &
0 &
0 \\
0 &
\mu_1 &
0 \\
0 &
0 &
\mu_3
\end{array}
\right).
\end{equation}
This is, by a basis change, the CP3 model~\eqref{CP3}.
We thus confirm $U(1) \oplus \Pi_2 = \text{CP3}$ in a certain basis.

\item Applying simultaneously $Z_2$~\eqref{Z_2} and $\Pi_2$~\eqref{Pi2} we
get immediately CP2~\eqref{CP2}. That is we confirm
$Z_2 \oplus \Pi_2 = \text{CP2}$ in a certain basis.
\end{itemize}

In a similar way we can show that $Z_2 \oplus \text{CP1} = \text{CP2}$,
$U(1) \oplus \text{CP1} = \text{CP3}$,
$U(2) \oplus \text{CP2}=\text{CP3}$,
$U(1) \oplus U(1)' =U(2)$, and
$\text{CP3} \oplus \text{CP3}' = U(2)$
in certain bases.

\end{document}